\documentclass[nobibnotes,superscriptaddress,showpacs,preprintnumbers,twocolumn,nofootinbib]{revtex4}
\usepackage{epsfig}
\begin{document}

\title{\mbox{}\\[10pt]
 A direct link between neutrinoless
double beta decay and leptogenesis\\ in a seesaw model with $S_4$ symmetry}

\author{Y. H. Ahn}
\email{yhahn@phys.sinica.edu.tw}
\affiliation{Institute of Physics, Academia Sinica, Taipei 115, Taiwan}
\author{Sin Kyu Kang}
\email{skkang@snut.ac.kr}
\affiliation{School of Liberal Arts, Seoul National Univ. of Technology, Seoul 139-743, Korea}
\author{C. S. Kim}
\email{cskim@yonsei.ac.kr, Corresponding Author}
\affiliation{Department of Physics and IPAP, Yonsei University, Seoul 120-749, Korea}
\author{T. Phong Nguyen}
\email{thanhphong@ctu.edu.vn}
\affiliation{Department of Physics and IPAP, Yonsei University, Seoul 120-749, Korea}
\affiliation{Department of Physics, Cantho University, Cantho, Vietnam}

\date{\today}

\begin{abstract}
\noindent We study how leptogenesis can be implemented in a seesaw model with $S_4$ flavor symmetry,
which leads to the neutrino tri-bimaximal mixing matrix
and degenerate right-handed (RH) neutrino spectrum.
Introducing a tiny soft $S_4$ symmetry breaking term in the RH neutrino mass matrix, we show that the
flavored resonant leptogenesis can be successfully realized,
which can lower the seesaw scale much so as to make it possible
to probe in colliders.
Even though such a tiny soft breaking term is essential for leptogenesis,
it does not significantly affect the low energy observables.
We also investigate how the effective light neutrino mass $|\langle m_{ee}\rangle |$
associated with neutrinoless double beta decay can be predicted
along with the neutrino mass hierarchies by imposing experimental data of low-energy observables.
We find a direct link between leptogenesis and  neutrinoless double beta decay characterized
by $|\langle m_{ee}\rangle|$
through a high energy CP phase $\phi$, which is correlated with low energy Majorana CP phases.
It is shown that our predictions of $|\langle m_{ee}\rangle|$
for some fixed parameters of high energy physics
can be constrained by the current observation of baryon asymmetry.
\end{abstract}

\maketitle %

\section{Introduction}

Recent experiments of the neutrino oscillation go into a new phase of precise determination
of mixing angles and mass squared differences
\cite{LowE data}, indicating that the tri-bimaximal (TBM) mixing for three flavors can be
regarded as so-called PMNS mixing matrix
$U_{\rm PMNS}\equiv U_{\rm TB}P_{\nu}$ in the lepton sector~\cite{TBM}
\begin{eqnarray}
 \label{TBM matrix 1}
 U_{\rm TB} = {\left(\begin{array}{ccc}
 \frac{\sqrt{2}}{\sqrt{3}} &\frac{1}{\sqrt{3}} & 0\\
 -\frac{1}{\sqrt{6}}&\frac{1}{\sqrt{3}} & \frac{1}{\sqrt{2}} \\
 -\frac{1}{\sqrt{6}} & \frac{1}{\sqrt{3}} & -\frac{1}{\sqrt{2}}
 \end{array}\right)}~, \end{eqnarray}
and $P_{\nu}$ is a diagonal matrix of phase for Majorana neutrinos.
However, properties related to the leptonic CP violation are completely
unknown yet.
The large mixing angles, which may be suggestive of a flavor symmetry,
are completely different from the quark mixing ones.
In last few years there have been lots of efforts in searching for
models which produce the TBM pattern for the neutrino mixing matrix, and
a fascinating way seems to be the use of some discrete non-Abelian flavor
groups added to the gauge groups of the Standard Model.
There is a series of models based on the symmetry group
$A_4$ \cite{A4}, $T'$ \cite{T'} and more recently $S_4$ \cite{old S4, new S4, S4}.

In addition to the explanation for the smallness of neutrino masses,
type-I seesaw model \cite{seesaw}, in which heavy right-handed singlet Majorana neutrinos
are introduced, has another appealing feature so-called
leptogenesis mechanism for the generation of the observed baryon
asymmetry of the Universe (BAU) through the decay of heavy Majorana neutrinos~\cite{review}.
If this BAU originated from leptogenesis, then CP symmetry in the leptonic sector must be broken.
So any observation of the leptonic CP violation, or demonstrating that CP is
not a good symmetry of the leptons, can strengthen our belief in leptogenesis.
For Majorana neutrinos there are two additional phases in $U_{\rm PMNS}$,
 one (or a combination) of which in principle can be explored through neutrinoless double beta ($0\nu2\beta$)
 decay~\cite{neutrinocp}.
Although the exact TBM mixing pattern forbids low energy CP violation measurable
in neutrino oscillation due to $U_{e3}=0$,
it may allow CP violation due to the Majorana phases.
Therefore, it is interesting to explore the existence of CP violation in the lepton sector
due to the Majorana CP phases in the light of leptogenesis. Since  $|\langle m_{ee}\rangle|$
depends on the Majorana CP phases, we examine if there exists a link between $0\nu2\beta$ decay and BAU.

In the neutrino models with some discrete flavor symmetries, it is worthwhile to examine
if leptogenesis can work out while keeping TBM pattern for neutrino mixing matrix.
Motivated by this issue, in this letter, we study how leptogenesis can work
in a seesaw model with $S_4$ flavor symmetry \cite{footnote1}. As anticipated,
leptogenesis can not work in seesaw models with exact $SU(2)_L\times U(1)_Y\times S_4$
symmetry mainly due to the fact that the discrete symmetry $S_4$ leads to zero value of $U_{e3}$
and thus could not generate lepton asymmetry.
To make leptogenesis successfully  realized, it is essential to break $S_4$ symmetry.
In the case that the right-handed (RH) neutrinos are hierarchical in mass, successful
leptogenesis requires that the RH neutrinos are superheavy, which makes inaccessible
to colliders and gives rise to the overproduction of gravitinos during reheating
in the supersymmetric scenarios \cite{moroi}. These problems can be avoided
if  we consider an almost degenerate heavy RH neutrino mass spectrum so that
lepton asymmetry can be resonantly generated \cite{resonant leptogenesis 1, mohapatra}.
In this paper, we especially study how this so-called resonant leptogenesis can be implemented
in a seesaw model with $S_4$ flavor symmetry \cite{ding2}
In this respect, the model has to produce degenerate RH neutrino mass spectrum at the leading order.
The $S_4$ seesaw model relevant to our purpose has been proposed in \cite{S4}.
We examine how resonant leptogenesis can be implemented by introducing
a small perturbation in heavy RH neutrino mass matrix $M_R$
while keeping Dirac neutrino mass matrix and charged lepton mass matrix unchanged
in the model \cite{S4}.
We also show that leptogenesis can be linked to the neutrinoless double beta
decay through seesaw mechanism with a small perturbation in $M_{R}$.

This work is organized as follows.
In Sec. II, we study how  low energy neutrino oscillation observables are predicted
in a supersymmetric seesaw model based on the flavor symmetry group $S_4$.
We also discuss about the effective neutrino mass associated with neutrinoless double beta decay.
In Sec. III, we examine how successful leptogenesis can be implemented
by introducing a soft symmetry breaking term in the heavy RH neutrino
mass matrix and show how leptogenesis can be liked to $0\nu2\beta$ decay.
Sec.~IV is devoted to our conclusion.
\section{Low energy observables}

Although there have been several proposals to construct lepton mass matrices
in the framework of seesaw incorporating $S_4$ symmetry \cite{new S4, ema},
in this paper, we consider the model proposed in \cite{S4},
which  gives rise to TBM mixing pattern of the lepton mixing matrix \cite{TBM}
and leads to degenerate heavy RH neutrino mass spectrum which is essential for resonant leptogenesis.
The model is supersymmetric and based on the flavor discrete group
$G_f = S_4\times Z_3\times Z_4$, where the three factors play different roles.
The reasons, why we consider superymmetry, are  to simplify the choice of
desirable vacuum alignment, and low scale leptogenesis is well motivated in supersymmetry
thanks to the gravitino problem. In general, there exists a contribution to leptogenesis via scalar right-handed
sneutrinos, but their effect depends on soft susy breaking terms. In our study, we do not consider
the contribution by simply assuming that the values of the soft susy breaking parameters do not give rise
to the contribution to leptogenesis.

The $S_4$ component controls the mixing angles, the auxiliary $Z_3$ symmetry
guarantees the misalignment in flavor space between the neutrino and
the charged lepton mass eigenstates, and the $Z_4$ component is crucial
to eliminating the unwanted couplings and reproducing the observed mass hierarchy.
In this framework the mass hierarchies are controlled
by the spontaneous breakdown of the flavor symmetry instead of
the Froggatt-Nielsen mechanism\cite{FN field}.
$S_4$ is the discrete group given by the permutations of four objects and
has been studied in literature \cite{old S4}, but with different aims and different results.
It is composed by 24 elements, divided into 5 irreducible representations:
two singlets, $1_1$ and $1_2$, one doublet, 2, and two triplets, $3_1$ and $3_2$.
The technical details of the group are shown in \cite{S4}.
The multiplication rules between the various representations are as follows;
$1_i\otimes1_j =1_{((i+j)\;{\rm mod}\;2)+1}$, $1_i\otimes2=2$,
 $1_i\otimes3_j =3_{((i+j)\;{\rm mod}\;2)+1}$, $2\otimes2=1_1\oplus1_2\oplus2$,$2\otimes3_i=3_1\oplus3_2$, $3_i\otimes3_i=1_1\oplus2\oplus3_1\oplus3_2$,
$3_1\otimes3_2=1_2\oplus2\oplus3_2\oplus3_2$, with $i,j=1,2$.
The matter fields in the lepton sector and the flavons under $G_f$ of the model are assigned as Table
\ref{particle content}.
\begin{widetext}
\begin{center}
\begin{table}[h]
\caption{\label{particle content} Representations of the matter fields in the lepton sector and
the flavons under $S_4 \times Z_{3} \times Z_4$.
And $\omega$ is the third root of unity, i.e. $\omega=e^{i2\pi/3}$.}
\begin{ruledtabular}
\begin{tabular}{ccccccccccccc}
Field &$l$&$e^{c}$&$\mu^{c}$&$\tau^{c}$&$\nu^{c}$&$h_{u,d}$&$\varphi$&$\chi$&$\vartheta$&$\eta$&$\phi$&$\Delta$\\
\hline
$A_4$&$\mathbf{3}_{1}$&$\mathbf{1}_{1}$&$\mathbf{1}_{2}$&$\mathbf{1}_{1}$&$\mathbf{3}_{1}$&$\mathbf{1}_{1}$&
$\mathbf{3}_{1}$&$\mathbf{3}_{2}$&$\mathbf{1}_{2}$&$\mathbf{2}$&$\mathbf{3}_{1}$&$\mathbf{1}_{2}$\\
$Z_3$&$\omega$&$\omega^{2}$&$\omega^{2}$&$\omega^{2}$&$1$&$1$&$1$&$1$&$1$&$\omega^{2}$&$\omega^{2}$&$\omega^{2}$\\
$Z_4$&$1$&$i$&$-1$&$-i$&$1$&$1$&$i$&$i$&$1$&1&1&$-1$\\
\end{tabular}
\end{ruledtabular}
\end{table}
\end{center}
\end{widetext}
The superpotential of the model in the lepton sector reads as follows
\begin{widetext}
\begin{eqnarray}
\label{lagrangian}
    w_l &=& \frac{y_{e1}}{\Lambda^3}e^c(l\varphi)_{1_1}(\varphi\varphi)_{1_1}h_d+\frac{y_{e2}}
    {\Lambda^3}e^c((l\varphi)_{2}(\varphi\varphi)_2)_{1_1}h_d +\frac{y_{e3}}{\Lambda^3}e^c((l\varphi)_{3_1}(\varphi\varphi)_{3_1})_{1_1}h_d
    + \frac{y_{e4}}{\Lambda^3}e^c((l\chi)_{2}(\chi\chi)_{2})_{1_1}h_d \nonumber\\
    &+&\frac{y_{e5}}{\Lambda^3}e^c((l\chi)_{3_1}(\chi\chi)_{3_1})_{1_1}h_d +\frac{y_{e6}}{\Lambda^3}e^c(l\varphi)_{1_1}(\chi\chi)_{1_1}h_d
    + \frac{y_{e7}}{\Lambda^3}e^c((l\varphi)_{2}(\chi\chi)_{2})_{1_1}h_d+\frac{y_{e8}}{\Lambda^3}e^c((l\varphi)_{3_1}(\chi\chi)_{3_1})_{1_1}h_d
    \nonumber\\
    &+& \frac{y_{e9}}{\Lambda^3}e^c((l\chi)_{2}(\varphi\varphi)_{2})_{1_1}h_d
    +\frac{y_{e10}}{\Lambda^3}e^c((l\chi)_{3_1}(\varphi\varphi)_{3_1})_{1_1} h_d+\frac{y_\mu}{2\Lambda^2}\mu^c(l(\varphi\chi)_{3_2})_{1_2}h_d
                +\frac{y_\tau}{\Lambda}\tau^c(l\varphi)_{1_1}h_d+...\\
    w_\nu     &=& \frac{y_{\nu 1}}{\Lambda}((\nu^cl)_2\eta)_{1_1} h_u + \frac{y_{\nu 2}}{\Lambda}((\nu^cl)_{3_1}\phi)_{1_1} h_u
    +\frac{1}{2}M(\nu^c\nu^c)_{1_1}+...,
\end{eqnarray}
\end{widetext}
where the dots denote higher order contributions.

The vacuum configuration can be determined from the vanishing of the derivatives of the
superpotentials $ w_l$  and $w_\nu$ with respect to each component of the driving fields, as shown in \cite{S4}.
Using this way, we can obtain the alignment of the VEVs of flavons as follows;
\begin{eqnarray}
\label{VEVs}
   \langle\varphi\rangle &=& (0,\upsilon_{\varphi},0)~,~\langle\chi\rangle=(0,\upsilon_{\chi},0)~,
   ~~~~\langle\vartheta\rangle=\upsilon_{\vartheta}~,
   \nonumber\\
   \langle\eta\rangle &=& (\upsilon_{\eta},\upsilon_{\eta})~,~~~\langle\phi\rangle=
   (\upsilon_{\phi},\upsilon_{\phi},\upsilon_{\phi})~,~\langle\Delta\rangle=\upsilon_{\Delta}~.
\end{eqnarray}
With these VEVs alignments as well as breaking of electroweak gauge symmetry,
the charged-lepton mass matrix is explicitly expressed as
\begin{eqnarray}
\label{charged lepton mass}
 m_l = {\rm Diag.}\Big( y_e\frac{\upsilon_\varphi^3}{\Lambda^3},
 y_\mu \frac{\upsilon_\varphi \upsilon_\chi}{\Lambda^2},
 y_\tau \frac{\upsilon_\varphi}{\Lambda}\Big) \upsilon_d~,
\end{eqnarray}
where we assume all components are real, and the neutrino sector gives rise to the following Dirac and
Majorana matrices
\begin{eqnarray}
\label{neutrino sector}
 m_\nu^d &=& {\left(\begin{array}{ccc}
 2be^{i\alpha_2} & ae^{i\alpha_1}-be^{i\alpha_2} & ae^{i\alpha_1}-be^{i\alpha_2}\\
 ae^{i\alpha_1}-be^{i\alpha_2} & ae^{i\alpha_1}+2be^{i\alpha_2} & -be^{i\alpha_2} \\
 ae^{i\alpha_1}-be^{i\alpha_2} & -be^{i\alpha_2} & ae^{i\alpha_1}+2be^{i\alpha_2}\end{array}\right)}
 \upsilon_u\nonumber\\
  &=& e^{i\alpha_1}{\left(\begin{array}{ccc}
 2be^{i\phi} & a-be^{i\phi} & a-be^{i\phi}\\
 a-be^{i\phi} & a+2be^{i\phi} & -be^{i\phi} \\
 a-be^{i\phi} & -be^{i\phi} & a+2be^{i\phi}\end{array}\right)} \upsilon_u~,\\
 M_R &=& {\left(\begin{array}{ccc}
 M  & 0  &  0\\
 0  &   0 &  M \\
 0  &   M  & 0 \end{array}\right)}~,
\end{eqnarray}
where we assume the quantity $M$, $a$ and $b$ are real and positive quantities,
and the relative phase $\phi\equiv\alpha_2-\alpha_1$ is
 the only physical phase because the phase $\alpha_1$ can be rotated away.
\begin{figure}[t]
\begin{tabular}{c}
\includegraphics[width=7.2cm]{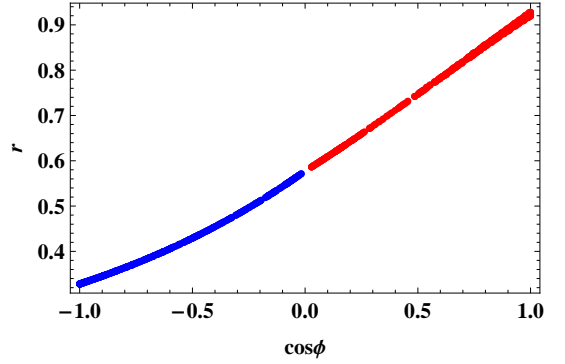}
\end{tabular}
\caption{\label{Fig1} Allowed parameter region of the ratio $r=b/a$ as a function of $\cos\phi$
constrained by the $1\sigma$ experimental data in Eq.~(\ref{LowE data}).
The thickness of the line reflects 1$\sigma$ uncertainty due to experimental data.
Here, the blue (dark) and red (light) curves correspond to the inverted
and normal mass ordering of light neutrino, respectively.}
\end{figure}
After seesawing, the effective light neutrino mass matrix is obtained
from seesaw formula $m_{\rm eff}=-(m_\nu^d)^TM_R^{-1}m_\nu^d$,
which can be diagonalized by the TBM mixing matrix as follows;
\begin{eqnarray}
\label{diagonalize mnu}
    U_{\nu}^Tm_{\rm eff}U_{\nu} &=& {\rm Diag.}(m_1, m_2, m_3),
\end{eqnarray}
where the mass eigenvalues are given as
\begin{eqnarray}
\label{light neutrino mass}
    m_{1} &=& m_{0}(1+9r^2-6r\cos\phi)~,\nonumber\\
    m_{2} &=& 4m_{0}~,\nonumber\\
    m_{3} &=& m_{0}(1+9r^2+6r\cos\phi)~,  \label{m_3}
\end{eqnarray}
with $r=b/a$, and $m_{0}=\upsilon^{2}_{u}a^{2}/M$ where $\upsilon_{u}=\upsilon\sin\beta$ and
$\upsilon=176$ GeV. And the lepton mixing
at low energy $U_{\rm PMNS}= U_\nu$ is given by
\begin{eqnarray}
\label{PMNS matrix}
    U_{\rm PMNS} = e^{-i\gamma_1/2} U_{\rm TB}{\rm Diag.}(1,e^{i\beta_1},e^{i\beta_2}),
\end{eqnarray}
where $\beta_1=\gamma_1/2,~\beta_2=(\gamma_1-\gamma_2)/2$ are the Majorana CP-violating phases, with
\begin{eqnarray}
    \gamma_1 &=& {\rm arg}\{(a-3be^{i\phi})^2\}~,\nonumber\\
    \gamma_2 &=& {\rm arg}\{-(a+3be^{i\phi})^2\}~.
\label{MjPhase}
\end{eqnarray}
The phase factored out to the left have no physical meaning,
since it can be eliminated by a redefinition of the charged lepton fields.
There are the nine physical quantities consisting of the three light neutrino masses,
the three mixing angles and the three CP-violating phases.
The mixing angles are entirely fixed by the $G_f$ symmetry group,
predicting TBM and in turn no Dirac CP-violating phase,
and the remaining 5 physical quantities $\beta_1, \beta_2, m_1, m_2$ and $m_3$,
are determined in terms of the five real parameters $M,\upsilon_u, a, b$ and $\phi$.
\begin{figure*}[t]
\begin{tabular}{cc}
\includegraphics[width=7.2cm]{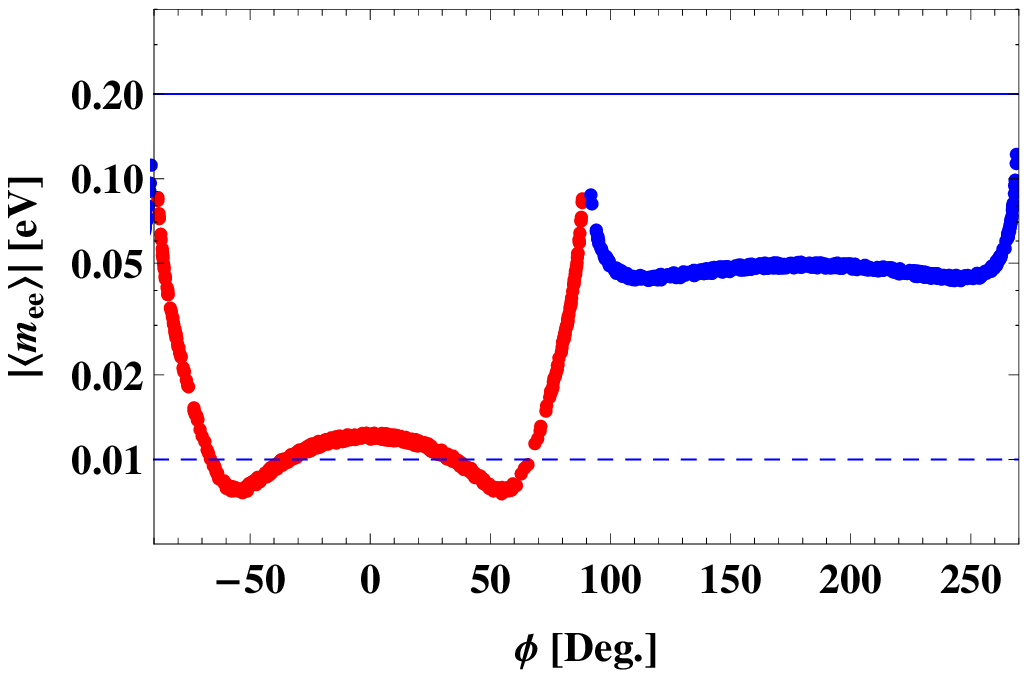}&
\includegraphics[width=7.2cm]{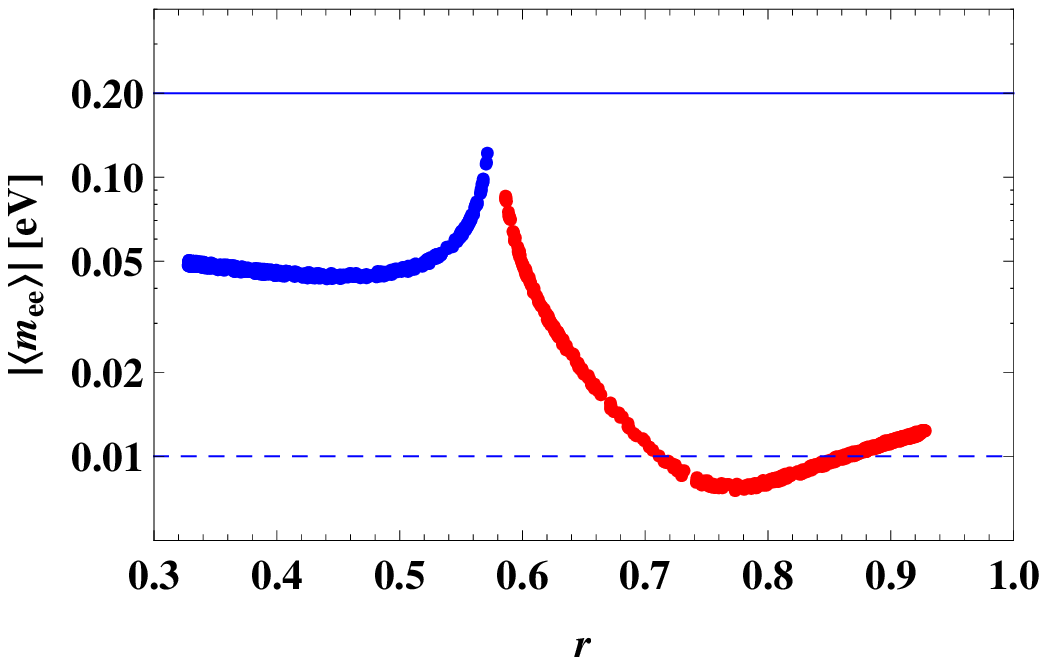}
\end{tabular}
\caption{\label{Fig2} Predictions of the effective mass $|\langle m_{ee}\rangle|$
for $0\nu 2\beta$ as a function of the phase $\phi$ in the left panel
and the ratio $r$ in the right panel based on the $1\sigma$ experimental results given
in Eq.~(\ref{LowE data}).
Here, in both panels the red (light) and blue (dark) curves correspond to the normal mass
spectrum of light neutrino and the inverted one, respectively.}
\end{figure*}

Because of the requirement of MSW resonance for solar neutrinos implying $\Delta m^{2}_{21}>0$,
and the observed hierarchy $|\Delta m^{2}_{31}|\gg\Delta m^{2}_{21}$,
there are two possible neutrino mass ordering depending on the sign of $\cos\phi$:
(i) $m_{1}<m_{2}<m_{3}$ (normal mass ordering) which corresponds to $\cos\phi > 0$ and
(ii) $m_{3}<m_{1}<m_{2}$ (inverted mass ordering) which corresponds to $\cos\phi <0$.
The solar and atmospheric mass-squared differences, which are given by
\begin{eqnarray}
 \Delta m^{2}_{21}&=& 3m^{2}_{0}(1-3r^{2}+2r\cos\phi)(5+9r^{2}-6r\cos\phi)~,\nonumber\\
 |\Delta m^{2}_{31}|&=& 24m^{2}_{0}r|\cos\phi|(1+9r^{2})~,
 \label{deltam2}
\end{eqnarray}
are constrained by the neutrino oscillation experiments. Since neutrino oscillation data
indicate that $\Delta m^{2}_{21}$ is positive, $1-3r^{2}+2r\cos\phi>0$.
It is interesting to see how the ratio $\Delta m^{2}_{21}/\Delta m^{2}_{31}$
leads to the correlation between the paremeters $r$ and $\cos\phi$.
{}From Eq.(\ref{deltam2}), we see that the ratio $\Delta m^{2}_{21}/\Delta m^{2}_{31}$
is independent of $m_0$, and that $\cos\phi=0$ is not allowed, which is reflected in Fig. 1.
For our purpose, we consider the experimental data at $1\sigma$~\cite{LowE data}:
\begin{eqnarray}
\label{LowE data}
    |\Delta m^2_{31}| &=& (2.29-2.52)\times 10^{-3}{\rm eV^2}~,\nonumber\\
    \Delta m^2_{21} &=& (7.45-7.88)\times 10^{-5}{\rm eV^2}~.
\end{eqnarray}
Hereafter, we use the $1\sigma$ confidence level experimental values of
low energy observables for our numerical calculations.
Imposing the experimental results, we present the correlations between $r$ and $\cos\phi$
for normal mass hierarchy (red-plot) and inverted one (blue plot) in Fig.~\ref{Fig1}.
From Eqs. (\ref{deltam2},\ref{LowE data}), we can also determine the value of $m_0$
as a function of $\cos\phi$.
Thanks to the zero entry in $U_{\rm PMNS}$,  $m_3$ does not contribute to
the effective neutrino mass and thus only the phase $\beta_1$ can contribute
to the $0\nu 2\beta$ decay amplitude, which can be written as
 $|\langle m_{ee}\rangle|=|m_{1}|U_{e1}|^{2}+m_{2}|U_{e2}|^{2}+m_{3}|U_{e3}|^{2}|$,
 where $U_{ei}$ (i=1,2,3) are the components of $U_{\rm PMNS}$.
As a result, the effective mass governing the $0\nu 2\beta$ decay is given by~\cite{S4}
\begin{widetext}
\begin{eqnarray}
\label{mee 1}
    |\langle m_{ee}\rangle| &=& \frac{1}{3}|2m_1 e^{2i\beta_1}+m_2|\nonumber\\
    &=& 2m_{0}\sqrt{1-4r\cos\phi+2r^{2}(2+3\cos2\phi)-12r^3\cos\phi+9r^{4}}~.
\end{eqnarray}
\end{widetext}
The behavior of $|\langle m_{ee}\rangle|$ is plotted in Fig.~\ref{Fig2}
as a function of the phase $\phi$ (left panel) and the ratio $r$ (right panel).
Here we note that  $m_{0}$ is related with other parameters through the seesaw formula
$m_{0}=a^{2}v^{2}_{u}/M$. For our numerical analysis, we take
 $m_{0}=0.0035-0.04$, $m_{0}=0.012-0.05$ for normal and inverted orderings, respectively,
 which are consistent with the experimental results given in Eq. (\ref{LowE data}).
Numerically, our prediction is turned out to be
$0.0076 {\rm eV} \leq |\langle m_{ee}\rangle| \leq 0.10$ eV for the normal mass ordering (red-plot)
and $0.043 {\rm eV} \leq |\langle m_{ee}\rangle| \leq 0.12$ eV for the inverted mass ordering (blue-plot),
where the upper limits come from the cosmological bound on neutrino mass scale.
We note that the existence of the lower bound on the prediction of $m_{ee}$ in the case of the normal
mass ordering is due to  $|cos \phi|\leq 1$, which
is easily understood from Eq. (\ref{m_3}), as also discussed in \cite{S4}.
In Fig. \ref{Fig2}, the horizontal solid and dashed lines correspond to the current lower
bound sensitivity (0.2 eV) \cite{present NDBD} and the future lower bound sensitivity
($10^{-2}$ eV) \cite{future NDBD} of $0\nu 2\beta$ experiments, respectively.
 The thickness of the lines correspond to the $1\sigma$ allowed ranges due to experimental results.

\begin{figure}[t]
\begin{tabular}{c}
\includegraphics[width=7.2cm]{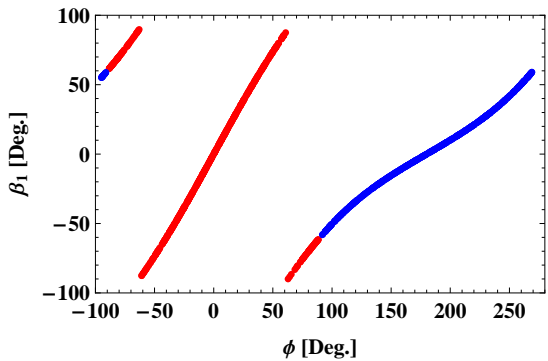}
\end{tabular}
\caption{\label{Fig2-1} Correlation of the Majorana CP phase $\beta_{1}$
with the phase $\phi$ constrained by the $1\sigma$ experimental
data in Eq.~(\ref{LowE data}). The red (light) and blue (dark) curves correspond
to the normal mass spectrum of light neutrino and the inverted one, respectively.}
\end{figure}

Using Eq.~(\ref{MjPhase}) we can obtain the explicit correlation between the phase $\phi$
and the Majorana phase $\beta_{1}$
\begin{eqnarray}\label{sinbeta1}
  \sin2\beta_{1}=\frac{6r\sin\phi(3r\cos\phi-1)}{1-6r\cos\phi+9r^{2}}~.
\end{eqnarray}
Note here that the size of $\tan2\beta_{1}$ is constrained by
Fig. \ref{Fig2-1} represents the correlation between the phase $\phi$ and
the Majorana phase $\beta_{1}$ for normal mass ordering (red-plot)
and inverted one (blue-plot).

\section{Leptogenesis through soft $S_4$ breaking}

In a basis where the charged current is flavor diagonal,
the diagonalization of $M_{R}$ leads to the mass eigenvalues of heavy RH neutrinos
given by
\begin{eqnarray}
  \label{VL VR}
  V_R^T M_RV_R={\rm Diag.}(M, M,-M)~,
\end{eqnarray}
where the diagonalizing matrix $V_{R}$ is
\begin{eqnarray}
  V_R = {\left(\begin{array}{ccc}
  1  & 0  &  0\\
  0  &  \frac{1}{\sqrt{2}} & \frac{-1}{\sqrt{2}} \\
  0  &  \frac{1}{\sqrt{2}} & \frac{1}{\sqrt{2}}
  \end{array}\right)}.\nonumber
\end{eqnarray}
After performing a basis rotation so that the heavy RH Majorana mass matrix $M_{R}$ becomes
diagonal by the unitary matrix $V_{R}$,
the Dirac mass matrix $m^{d}_{\nu}$ gets modified to
\begin{eqnarray}
 m_\nu^d\rightarrow Y_\nu\upsilon_u=V_R^T m_\nu^d
\end{eqnarray}
where the Yukawa coupling matrix $Y_{\nu}$, is given as
\begin{eqnarray}
    \label{Ynu}
    Y_\nu &=& {\left(\begin{array}{ccc}
    2be^{i\phi} & a-be^{i\phi}  &  a-be^{i\phi}\\
    \sqrt{2}(a-be^{i\phi}) & \frac{a+be^{i\phi}}{\sqrt{2}}  & \frac{a+be^{i\phi}}{\sqrt{2}} \\
    0  &   -\frac{a+ 3be^{i\phi}}{\sqrt{2}}  &  \frac{a+ 3be^{i\phi}}{\sqrt{2}} \end{array}\right)}~.
\end{eqnarray}
Here, we notice that the CP phase $\phi$ existing in $m^{d}_{\nu}$ obviously
gives rise to low-energy CP violation.
On the other hand, leptogenesis is associated with both $Y_{\nu}$ itself
and the combination of Yukawa coupling matrix, $H\equiv Y_{\nu}Y^{\dag}_{\nu}$,
which is given as
\begin{widetext}
\begin{eqnarray}
  H &= & {\left(\begin{array}{ccc}
  2a^2+6b^2-4ab\cos\phi  & \sqrt{2}(a^2-3b^2+2ab\cos\phi)  &  0\\
  \sqrt{2}(a^2-3b^2+2ab\cos\phi) & 3a^2+3b^2-2ab\cos\phi &  0 \\
  0  &   0  & a^2+9b^2+6ab\cos\phi \end{array}\right)}~.
  \label{H matrix}
\end{eqnarray}
\end{widetext}
We see that all ${\rm Im}[H_{ij}]$ vanish and thus leptogenesis could not take place.
However, when we consider low scale much below $10^9$ GeV to avoid the abundance of
gravitino overproduction problem~\cite{Ellis:1984eq} within a super-symmetric version,
the $\tau$ and $\mu$ Yukawa couplings comes into equilibrium as the lepton asymmetry
is being created. In that case, lepton number asymmetries and washout effects become flavor
dependent, and this can give rise to a final baryon asymmetry which is different in size
from the one generated if flavor effects are ignored ~\cite{flavored, Abada:2006fw}.
We call it flavored leptogenesis  that occurs in general for
$T\sim M_{i}(1+\tan^{2}\beta)\lesssim10^{9}$ GeV in supersymmetric seesaw model~\cite{Abada:2006fw}.
In addition, since exact degenerate mass spectrum of the heavy RH neutrinos prevents leptogenesis to be
occurred, in order for flavored leptogenesis to be viable,
the degeneracy of the heavy RH neutrino masses should be lifted.

For an almost degenerate heavy Majorana neutrino mass spectrum, leptogenesis
can be naturally implemented through the resonant-leptogenesis
framework \cite{resonant leptogenesis 1, mohapatra, resonant leptogenesis 2}.
In this case, the CP asymmetry generated by the \emph{i}-th heavy Majorana neutrino decaying
into a lepton flavor $\alpha$ is given by \cite{resonant leptogenesis 3}
\begin{eqnarray}
  \varepsilon_{i}^\alpha
  &=&\sum_{j\neq i} \frac{{\rm Im}\Big[H_{ij}(Y_\nu)_{i\alpha}(Y_\nu)_{j\alpha}^\ast\Big]}{16\pi H_{ii}\delta_N^{ij}}
  \Big(1+\frac{\Gamma_j^2}{4M_j\delta_N^{ij2}}\Big),
 \label{f-cpasym 0}
\end{eqnarray}
where $\Gamma_j=H_{jj}M_j/8\pi$ is the decay width of the \emph{j}-th
right-handed Majorana neutrino and $\delta_N^{ij}$ is
the mass splitting parameter defined as
\begin{eqnarray}
  \delta_N^{ij}= 1-\frac{M_j}{M_i}.
 \label{mass slitting}
\end{eqnarray}

In order for resonant leptogenesis to be implemented successfully,
soft terms of the form $\epsilon M\overline{\nu_i^c}\nu_j^c~$
with small dimensionless parameter $\epsilon$ \cite{soft breaking} are added  in Eq.~(3),
which lift the degeneracy of the heavy Majorana neutrino masses.
Although there are several possibilities to incorporate the breaking parameter $\epsilon$ in $M_{R}$,
which lead to the mass slitting parameter $|\delta_N^{ij}|\sim \epsilon$.
Without loss of generality, we introduce a breaking term of the form
$\epsilon M\overline{\nu_2^c}\nu_2^c$, which modifies the RH neutrino mass matrix as
\begin{eqnarray}
\label{Soft braking MR}
     M_R= {\left(\begin{array}{ccc}
     M  & 0  &  0\\
     0  &   \epsilon M &  M \\
     0  &   M  & 0 \end{array}\right)},
\end{eqnarray}
where the parameter $\epsilon$ is assumed to be real.
$M_R$ is diagonalized as $\tilde{V}_R^TM_R\tilde{V}_R = {\rm Diag.}(M_1, M_2,M_3)$
with real eigenvalues given as
\begin{eqnarray}
  \label{Modified Mi}
  M_1= M~,~M_2\simeq M\Big(1+\frac{\epsilon}{2}\Big)~,~M_3\simeq -M\Big(1-\frac{\epsilon}{2}\Big)~
\end{eqnarray}
and the diagonalizing matrix $\tilde{V}_{R}$ is written as
\begin{eqnarray}
    \label{Modified VR}
    \tilde{V}_R\simeq {\left(\begin{array}{ccc}
    1  &    0    &  0 \\
    0  &  \frac{1}{\sqrt{2}}+ \frac{\epsilon}{4\sqrt{2}}& \frac{-1}{\sqrt{2}}+ \frac{\epsilon}{4\sqrt{2}} \\
    0  &  \frac{1}{\sqrt{2}}- \frac{\epsilon}{4\sqrt{2}} & \frac{1}{\sqrt{2}}+ \frac{\epsilon}{4\sqrt{2}} \end{array}\right)}~.
\end{eqnarray}
In the basis where the heavy RH mass matrix real and diagonal,
the Dirac Yukawa coupling matrix now reads as
\begin{eqnarray}
\label{modified Ynu}
  \tilde{Y}_\nu &=& \frac{1}{\upsilon_u}\tilde{V}_R^Tm_\nu^d,
\end{eqnarray}
where $m_\nu^d$ and $\tilde{V}_R$ are given in Eq.~(\ref{neutrino sector})
and Eq.~(\ref{Modified VR}), respectively.
Assuming $\epsilon\ll1$, we find that the matrix $\tilde{Y}_\nu\tilde{Y}_\nu^\dag$ is
real and almost the same as $\tilde{Y}_\nu\tilde{Y}^{\dag}_\nu\simeq H$ given in Eq.~(\ref{H matrix}).
As a result, the contributions of $N_{3}$ to lepton asymmetries $\varepsilon^{\alpha}_{i}$
can be negligible, due to $H_{12(21)}\gg H_{13(31)}\simeq H_{23(32)}\simeq0$.
From the heavy Majorana neutrino masses given in Eq.~(\ref{Modified Mi})
we can obtain the mass splitting parameter as follows,
\begin{eqnarray}
\label{mass slitting1}
 \delta_N^{21}=-\delta_N^{12}\simeq \frac{\epsilon}{2}~.
\end{eqnarray}
Then, combining with Eqs.~(\ref{Ynu}, \ref{H matrix}) and Eq.~(\ref{mass slitting1}),
the flavor dependent CP asymmetries $\varepsilon_i^\alpha$
can be obtained as follows
\begin{widetext}
\begin{eqnarray}
\label{leptonasym}
 \varepsilon^{e}_{1} &\simeq& -\frac{a^{2}(1+2r\cos\phi-3r^{2})}{\epsilon4\pi(1-2r\cos\phi+3r^2)}r\sin\phi~,~~~
 \varepsilon^{\mu}_{1}\simeq\varepsilon^{\tau}_{1}\simeq \frac{a^{2}(1+2r\cos\phi-3r^{2})}{\epsilon8\pi(1-2r\cos\phi+3r^2)}r\sin\phi\nonumber\\
 \varepsilon^{e}_{2} &\simeq& -\frac{a^{2}(1+2r\cos\phi-3r^{2})}{\epsilon2\pi(3-2r\cos\phi+3r^2)}r\sin\phi~,~~~
 \varepsilon^{\mu}_{2}\simeq\varepsilon^{\tau}_{2}\simeq \frac{a^{2}(1+2r\cos\phi-3r^{2})}{\epsilon4\pi(3-2r\cos\phi+3r^2)}r\sin\phi
\end{eqnarray}
\end{widetext}
In these expressions, the values of the parameters $\phi$ and $ r$ are determined from the analysis
as demonstrated in sec.II,
whereas $\epsilon$ and $a$ are arbitrary. However, since the seesaw relation
$a^{2}=m_{0}M/\upsilon^{2}_{u}$ as defined in Eq.~(\ref{light neutrino mass}),
the value of $a$ depends on the magnitude of $M$ once $m_{0}$ is determined.
Thus, in our numerical analysis, we take $M$ and $\epsilon$
as input in the estimation of lepton asymmetry.
Here, we note that although $\epsilon$ and $M$ are taken to be independent parameters in our analysis,
the predictions of the lepton asymmetries $\varepsilon^{\alpha}_{1,2}$
depends only on the ratio $M/\epsilon$. We see from Eq.~(\ref{leptonasym})
that the asymmetries can  be substantially enhanced by lowering $\epsilon$.
However, the value of $\epsilon$ is constrained by the validity of the perturbation.
In order for the perturbative approach to be valid, the tree-level decay width $\Gamma_{i}$
must be much smaller than the mass difference:
\begin{eqnarray}
\label{perturbative}
 \Gamma_{i}=\frac{[Y_{\nu}Y^{\dag}_{\nu}]_{ii}}{8\pi}M_{i}\ll M_{2}-M_{1}=\delta_{N}M_{2}~,~~i=1,2~.
\end{eqnarray}
It is approximately given by
\begin{eqnarray*}
\Gamma_{i}\simeq a^{2}(1+3r^{2}-2r\cos\phi)/4\pi\ll \delta_{N}^{21}.
\end{eqnarray*}
Due to $a^{2}=m_{0}M/\upsilon^{2}_{u}$ and $\delta_{N}^{21}\simeq\epsilon/2$,
if we take the seesaw scale, as an example, to be $M=10^{6}$ GeV,
the value of $a$ is the order of ${\cal O}(10^{-5})$,
which requires $\epsilon\gg10^{-10}$ for $\delta_{N}\simeq\epsilon/2$.
\begin{figure*}[t]
\begin{tabular}{cc}
\includegraphics[width=7.2cm]{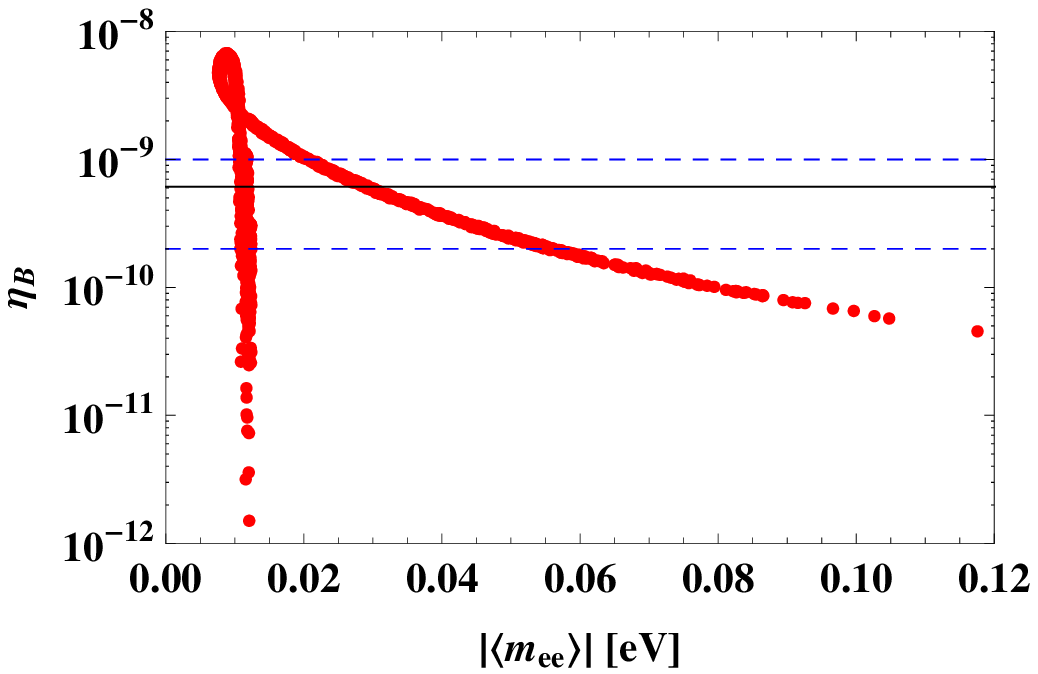}&
\includegraphics[width=7.2cm]{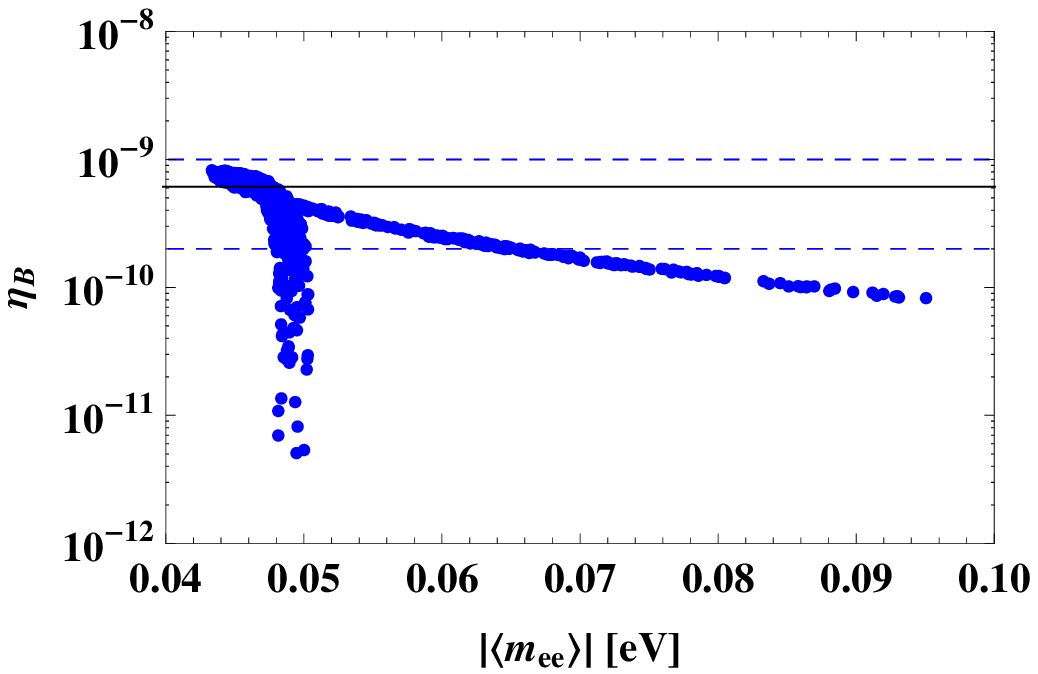}
\end{tabular}
\caption{\label{Fig3}
The prediction of $\eta_B$ as a function of $|\langle m_{ee}\rangle|$
for $M=10^6$ GeV and the soft breaking parameter $\epsilon =10^{-7}$.
The left-plot and the right-plot represent the normal mass ordering of light neutrino
and the inverted one, respectively.
The solid horizontal line and the dotted horizontal lines corresponds
to the experimental value of baryon asymmetry,
$\eta^{\rm CMB}_B=6.1\times 10^{-10}$, and phenomenologically allowed
regions $2\times10^{-10}\leq\eta_{B}\leq10^{-9}$, respectively.}
\end{figure*}

Once the initial values of $\varepsilon^{\alpha}_{i}$ are fixed, the final result of $\eta_{B}$
can be obtained by solving a set of flavor dependent
Boltzmann equations including the decay, inverse decay, and scattering processes as well
as the nonperturbative sphaleron interaction.
In order to estimate the wash-out effects, we introduce the parameters
$K^{\alpha}_{i}$ which are the wash-out factors due to the inverse decay of the
Majorana neutrino $N_{i}$ into the lepton flavor  $\alpha(=e,\mu,\tau)$ \cite{Antusch}.
The explicit form of $K^{\alpha}_{i}$ is given by
\begin{eqnarray}
   K^{\alpha}_{i}=\frac{\Gamma^{\alpha}_{i}}{H(M_{i})}=(Y^{\dag}_{\nu})_{\alpha i}(Y_{\nu})_{i\alpha}\frac{\upsilon^{2}_{u}}{m_{\ast}M_{i}}
 \label{washout01}
 \end{eqnarray}
where $\Gamma^{\alpha}_{i}$ is the partial decay width of $N_{i}$ into lepton flavor
$\alpha$ and Higgs scalars,
$H(M_{i})\simeq(4\pi^{3}g_{\ast}/45)^{\frac{1}{2}}M^{2}_{i}/M_{Pl}$
with the Planck mass $M_{Pl}=1.22\times10^{19}$ GeV
and the effective number of degrees of freedom $g_{\ast}\simeq228.75$ is
the Hubble parameter at temperature $T=M_{i}$,
and the equilibrium neutrino mass $m_{\ast}\simeq10^{-3}$. From Eqs.~(\ref{Ynu}) and ~(\ref{washout01})
we can obtain the washout parameters as follows
\begin{eqnarray}
 K^{e}_{1}&\simeq& 4r^{2}\frac{m_{0}}{m_{\ast}}~,\nonumber\\
 K^{\mu,\tau}_{1}&\simeq& \frac{m_{0}}{m_{\ast}}(1-2r\cos\phi+r^{2})~, \nonumber\\
 K^{e}_{2}&\simeq& \frac{2m_{0}}{m_{\ast}}(1-2r\cos\phi+r^{2})~,\nonumber\\
 K^{\mu,\tau}_{2}&\simeq& \frac{m_{0}}{2m_{\ast}}(1+2r\cos\phi+r^{2})~.
\label{washout02}
\end{eqnarray}
Since we take $M$ to be $10^6$ GeV, each lepton asymmetry for a single flavor in
Eq.~(\ref{leptonasym}) is weighted differently
by the corresponding washout parameter given by Eq.~(\ref{washout02}),
and appears with different weight in the final formula
for the baryon asymmetry~\cite{Abada:2006fw,Antusch} as follows;
\begin{widetext}
\begin{eqnarray}
   \eta_B \simeq-10^{-2}\sum_{N_{i}}\Big[\varepsilon^{e}_{i}\kappa\Big(\frac{93}{110}K^{e}_{i}\Big)
   +\varepsilon^{\mu}_{i}\kappa\Big(\frac{19}{30}K^{e}_{i}\Big)+\varepsilon^{\tau}_{i}\kappa\Big(\frac{19}{30}K^{e}_{i}\Big)\Big]~,
 \end{eqnarray}
\end{widetext}
with wash-out factor
\begin{eqnarray}
\label{washout}
 \kappa^{\alpha}_{i}\simeq\Big(\frac{8.25}{K^{\alpha}_{i}}+\Big(\frac{K^{\alpha}_{i}}{0.2}\Big)^{1.16}\Big)^{-1}~.
\end{eqnarray}
As can be seen from Eqs.~(\ref{leptonasym},\ref{washout02}),
since the lepton asymmetries in $\mu$ and $\tau$ flavors are equal to the first order,
satisfying $\varepsilon^{\mu}_{1(2)}+\varepsilon^{\tau}_{1(2)}=-\varepsilon^{e}_{1(2)}$,
and the washout factors in $\mu$ and $\tau$ are also equal,
the value of baryon asymmetry can be obtained as
\begin{eqnarray}
   \eta_B \simeq 10^{-2}\{\varepsilon^{e}_{1}(\kappa^{\mu}_{1}-\kappa^{e}_{1})+\varepsilon^{e}_{2}(\kappa^{\mu}_{2}-\kappa^{e}_{2})\}~.
 \label{etaB}
\end{eqnarray}
Numerically, the washout factors for normal mass spectrum of neutrino and the inverted one are given as
\begin{eqnarray}
 |\kappa^{e}_{1}-\kappa^{\mu}_{1}| &\lesssim& 0.06,~~|\kappa^{\mu}_{2}-\kappa^{e}_{2}|\lesssim0.05~, \nonumber\\
 \kappa^{e}_{1}-\kappa^{\mu}_{1} &\lesssim& 0.02,~~\kappa^{\mu}_{2}-\kappa^{e}_{2}\lesssim 0.05~,
\label{kappa}
\end{eqnarray}
respectively. Thus, taking Eqs.~(\ref{etaB},\ref{kappa}) into account,
$|\varepsilon^{e}_{1(2)}|\sim10^{-6-7}$ is needed to obtain a successful leptogenesis,
which in turn means that the value of soft breaking parameter $\epsilon\lesssim10^{-6}$
is required for $M=10^{6}$ GeV.

Before going to discuss the value of $\eta_B$, we consider the effects of soft breaking
term on the light neutrino observables.
The effective neutrino mass matrix is modified due to the soft breaking term and
thus it can be diagonalized as
$\tilde{U}_{\nu}^T\tilde{m}_{\rm eff}\tilde{U}_{\nu} = {\rm Diag.}(\tilde{m}_1,\tilde{m}_2,\tilde{m}_3)$
with the real eigenvalues
\begin{eqnarray}
 \tilde{m}_{i} &=& m_i+m_i{\cal O}(\epsilon)~,
\label{light neutrino mas 1}
\end{eqnarray}
and the mixing matrix $\tilde{U}_{\nu}=U_{\rm PMNS}$ can be written as
\begin{eqnarray}
 U_{\rm PMNS}=U_{TB}P_{\nu}+\delta U(\epsilon)P_{\nu}
\label{Modified PMNS matrix}
\end{eqnarray}
where $P_{\nu}={\rm Diag.}(1,e^{i\beta_1},e^{i\beta_2})$ is diagonal matrix of
Majorana phases defined in Eq.~(\ref{PMNS matrix}).
Note here that, since $\epsilon$ is real, the Majorana CP phases $\beta_{1,2}$
are not affected from the soft breaking.
However, it is obvious that the effects of $\epsilon$ to the light neutrino mass eigenvalues
and the neutrino mixing angles are negligible
due to $\epsilon\lesssim10^{-6}$.
As expected, the value of  $|\langle m_{ee}\rangle|$ is almost the same as Eq.~(\ref{mee 1}).


The predictions for $\eta_B$ as a function of $|\langle m_{ee}\rangle|$ are shown
in Fig.~\ref{Fig3} where we have used $M=10^6$ GeV,
$\epsilon=10^{-7}$ and $\tan\beta=2.5$
 as inputs. Please note that if there is a mass splitting of Majorana neutrinos by
a renormalization group equation, the value of $\tan\beta$ can be crucial to have a successful leptogenesis.
However, its value does  not affect significantly our results. So,
for simplicity, we take $\tan\beta=2.5$.
The horizontal solid and dashed lines correspond to the central value of the experiment result of  BAU
$\eta^{\rm CMB}_B=6.1\times 10^{-10}$~\cite{cmb} and the phenomenologically allowed regions
$2\times10^{-10}\leq\eta_B\leq10^{-9}$, respectively.
As shown in Fig.~\ref{Fig3}, the current observation of $\eta^{\rm CMB}_{B}$
can narrowly constrain the value of $|\langle m_{ee}\rangle|$ for
the normal hierarchical mass spectrum of light neutrino and inverted one, respectively in the case
that the scale of leptogenesis is $10^{3} < M < 10^{6}$ GeV and $\epsilon$ is small enough.
Combining the results presented in Figs.~\ref{Fig2} and~\ref{Fig2-1} with those from leptogenesis,
we can pin down the Majorana CP phase $\beta_{1}$ via the parameter $\phi$.

\section{Conclusions}

We have examined how leptogenesis can be implemented in the seesaw model with $S_4$ flavor symmetry
which leads to the neutrino TBM  mixing matrix
and degenerate RH neutrino spectrum.
In order for leptogenesis to be viable, $Y_\nu Y_\nu^\dagger$ should contain nontrivial
imaginary part and
the degeneracy of heavy right-handed Majorana masses has to be lifted.
In our study, we have shown that flavored resonant leptogenesis can be successfully realized
by introducing a tiny soft $S_4$ symmetry breaking term
in $M_{R}$, which can much lower the seesaw scale.
Even though we can lower the seesaw scale down to TeV scale, it would be difficult to probe it directly
in collider experiments because the couplings between the heavy neutrinos and the light neutrinos are so small
in our scenario.
Even though such a tiny soft breaking term is essential for leptogenesis, it does not significantly
affect the low energy observables.
We have also investigated how the effective neutrino mass $|\langle m_{ee}\rangle |$ associated
with  $0\nu 2\beta$ decay can be predicted
along with the light neutrino mass hierarchies in our scenario for leptogenesis by imposing
experimental data of low-energy observables.
Interestingly enough, we have found a direct link between leptogenesis and neutrinoless double
beta decay characterized by $|\langle m_{ee}\rangle|$
through a high energy CP phase $\phi$ which is correlated with low energy Majorana CP phases.
We also have shown that our predictions of $|\langle m_{ee}\rangle|$  for some fixed parameters
can be constrained by the current observation of baryon asymmetry in the case
that the scale of leptogenesis is $10^{3} < M < 10^{6}$ GeV and $\epsilon$ is small enough.

\acknowledgments{
\noindent
YHA is supported by the National Science Council of R.O.C. under Grants No:
 NSC-97-2112-M-001-004-MY3.
The work of SKK was supported in part by the
Korea Research Foundation(KRF) grant funded by the Korea government(MEST) (2009-
0069755).
The work of CSK and TPN was supported in part by Basic Science Research Program through the NRF of Korea
funded by MOEST (2009-0088395) and  in part by KOSEF through the Joint Research Program
(F01-2009-000-10031-0).
 \\
}


\begin{thebibliography}{99}
\bibitem{ding2} The possibility of resonant leptogenesis in the model we consider has been mentioned
 in \cite{S4}.
\end{thebibliography}
\end{document}